\documentclass[epj,nopacs]{svjour}

\usepackage{graphicx}
\usepackage{fancyhdr}
\def\makeheadbox{{
 }}
\def\mytitle{My title} 
\def\myauthors{My name}  
\def\mytype{My type of session}
\def\mysession{My session}

\def\mytitle{Prospects for the Determination of Higgs Boson Properties at the LHC} 
\def\myauthors{Christoph Ruwiedel}    
\def\mytype{Contributed Talk}    
\def\mysession{Colliders - Higgs Phenomenology}

\begin{document}
\title{Prospects for the Determination of Higgs Boson Properties at the LHC}
\author{Christoph Ruwiedel, for the ATLAS and CMS collaborations
}                     
%
%
\institute{Physikalisches Institut, Universit\"at Bonn, Nussallee 12, 53115 Bonn, Germany}
%
\date{}

\abstract{The prospects for the determination of the properties of a hypothetical boson discovered at the
  Large Hadron Collider using measurements to be performed with the ATLAS and CMS detectors are
  summarized. The properties that are discussed are the mass of the boson, its width, the strength of its
  couplings to fermions, the strength and structure of its couplings to gauge bosons and its spin and CP
  quantum numbers.}

\maketitle

\section{Introduction}
\label{intro}
The determination and testing of the theory underlying electroweak symmetry breaking and the description of
particle masses is one of the main objectives of the ATLAS and CMS experiments at the Large Hadron Collider
(LHC). Monte Carlo studies indicate \cite{ATLAS_TDR,CMS_TDR} that the discovery of the Higgs boson associated
with the symmetry breaking mechanism of the Standard Model, if it exists, will be possible over the full mass
range that has not yet been excluded. Due to the variety of suggested extensions of the Standard Model and
alternative descriptions of electroweak symmetry breaking, however, the discovery of a particle by itself is
unlikely to be sufficient to uniquely identify the model implemented in nature. A systematic measurement of
the properties of a newly discovered particle will be important to establish its relation to symmetry
breaking, constrain model predictions and exclude alternative models as far as possible.

\section{Mass and width of the Higgs Boson}

\begin{figure*}
\begin{center}
\includegraphics[width=0.4\textwidth]{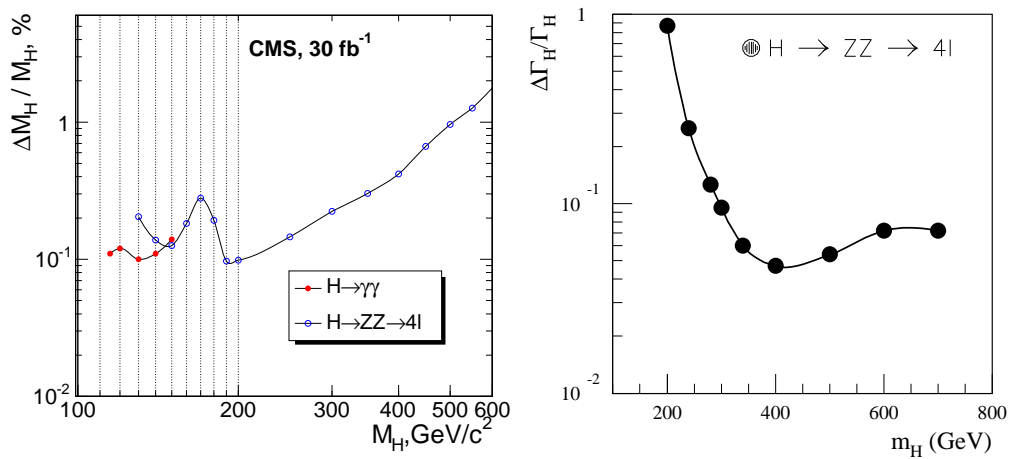}
\includegraphics[width=0.377\textwidth]{hwidth.epsi}
\end{center}
\caption{Expected precision, taking into account statistical errors only, of the measurement of the Higgs
  boson mass with the CMS experiment \cite{CMS_TDR} (\emph{left}) and expected precision for the measurement
  of the Higgs boson width with the ATLAS experiment for an integrated luminosity of 300\,fb$^{-1}$
  \cite{ATLAS_TDR} (\emph{right}).}
\label{fig:mass_width}
\end{figure*}

The mass of the Standard Model Higgs boson can be measured with a high precision in the decay channels $H \to
\gamma\gamma$ and $H \to ZZ \to 4l$ by direct reconstruction from the measured four-momenta of the final state
particles. The expected precision from a fit to the invariant mass distribution for the CMS experiment, taking
into account statistical errors only, is shown in Fig.\ref{fig:mass_width} (\emph{left}). An integrated
luminosity of 30$\,$fb$^{-1}$, which is expected to be obtained after the first three years of LHC operation,
is assumed for the plot. The expected precision is better than $0.3\%$ for Higgs boson masses smaller than
about $300\,$GeV. For the ATLAS experiment, assuming an integrated luminosity of 300$\,$fb$^{-1}$ and taking
into account also the systematic errors on the absolute energy scale, the precision is expected to be $0.1\%$
for Higgs boson masses smaller than $400\,$GeV. For this result the scale uncertainty for leptons and photons
is assumed to be $0.1\,\%$, which is considered a conservative estimate \cite{ATLAS_TDR}. For larger Higgs
boson masses the precision degrades to around $1\%$ at $600\,$GeV.

If the above decay channels turn out to be suppressed for the particle under study the measurement of the mass
will have a larger uncertainty. The invariant mass can be reconstructed in the channel $ttH \to ttbb$ directly
from the reconstructed final state, in the channel $H \to WW \to l\nu jj$ by using the mass of the $W$ boson
as a constraint and in the channel $H \to \tau\tau$ by using the approximation that the momenta of all decay
products of a $\tau$ lepton are collinear to the momentum of the $\tau$ lepton itself. Finally, in the channel
$H \to WW \to ll\nu\nu$ the mass can be extracted from a fit to the distribution of the transverse mass of the
lepton-neu\-tri\-no system.

The expected precision of a measurement of the Standard Model Higgs boson width for masses larger than
200\,GeV with the ATLAS experiment for an integrated luminosity of 300\,fb$^{-1}$ is shown in
Fig.\ref{fig:mass_width}. The precision is expected to be better than 8\% for masses larger than about
270\,GeV. For Higgs boson masses smaller than 200\,GeV the observed width of the Higgs resonance peak will be
dominated by the detector resolution and a direct measurement of the width will not be possible. Also, at the
LHC some decay channels of the Higgs boson such as $H \to gg$ and the decay to quarks lighter than $b$ will
not be observable and for some decay channels such as $H \to b\bar{b}$ the experimental uncertainties will be
large. Therefore, the width cannot be determined in a model-independent way from the observed decay
rates. However, an indirect determination using assumptions about the model can be performed in this mass
region, as discussed in Sec.\ref{Sec:Duehrssen_couplings}.

\section{Measurement of couplings of a CP even scalar boson}
\label{Sec:Duehrssen_couplings}

The prospects for the measurement of the couplings of a CP even scalar Higgs boson candidate to fermions and
gauge bosons were studied \cite{DUEHRSSEN_ATLAS,DUEHRSSEN_PRD} by performing a likelihood fit to the expected
numbers of events taken from ATLAS analyses for the channels listed in
Tab.\ref{tab:couplings_fit_channels}. Sources of systematic errors that were taken into account in the fit are
the luminosity, detector effects, the background normalization and theoretical uncertainties on the cross
sections. The event rate in a given channel can be approximated \cite{DUEHRSSEN_PRD} as
\begin{equation}
\sigma(H) \cdot \mathrm{BR}(H \to x) = \frac{\sigma(H)^{\mathrm{SM}}}{\mathrm{\Gamma}^{\mathrm{SM}}_{p}} \cdot
\frac{\mathrm{\Gamma}_{p}  \mathrm{\Gamma}_{x}}{\mathrm{\Gamma}}
\end{equation}
with the partial width $\mathrm{\Gamma}_{p}$ describing the production process and the partial width
$\mathrm{\Gamma}_{x}$ describing the decay.

\begin{table}
\caption{Production and decay channels and corresponding mass ranges used for the fit of fermion and gauge
  boson couplings.}
\label{tab:couplings_fit_channels}
\begin{tabular}{lll}
\hline\noalign{\smallskip}
\multicolumn{1}{c}{Production} & \multicolumn{1}{c}{Decay} & \multicolumn{1}{c}{Mass range}  \\
\noalign{\smallskip}\hline\noalign{\smallskip}
Gluon & $H \to ZZ^{(*)} \to 4l$ & (110 - 200)\,GeV \\
Fusion & $H \to WW^{(*)} \to l\nu l\nu$ & (110 - 200)\,GeV \\
& $H \to \gamma \gamma$ & (110 - 150)\,GeV \\
\noalign{\smallskip}\hline\noalign{\smallskip}
Weak Boson & $H \to ZZ^{(*)} \to 4l$ & (110 - 200)\,GeV \\
Fusion & $H \to WW^{(*)} \to l\nu l\nu$ & (110 - 190)\,GeV \\
& $H \to \tau\tau \to l\nu\nu l\nu\nu$ & (110 - 150)\,GeV \\
& $H \to \tau\tau \to l\nu\nu$had$\nu$ & (110 - 150)\,GeV \\
& $H \to \gamma \gamma$ & (110 - 150)\,GeV \\
\noalign{\smallskip}\hline\noalign{\smallskip}
$t\bar{t}H$ & $H \to WW^{(*)}$ & (120 - 200)\,GeV \\
& $\quad\, \to l\nu l\nu (l\nu)$ & \\
& $H \to b\bar{b}$ & (110 - 140)\,GeV \\
& $H \to \tau\tau$ & (110 - 120)\,GeV \\
\noalign{\smallskip}\hline\noalign{\smallskip}
$WH$ & $H \to WW^{(*)}$ & (150 - 190)\,GeV \\
& $\quad\, \to l\nu l\nu (l\nu)$ & \\
& $H \to \gamma \gamma$ & (110 - 120)\,GeV \\
\noalign{\smallskip}\hline\noalign{\smallskip}
$ZH$ & $H \to \gamma \gamma$ & (110 - 120)\,GeV \\
\noalign{\smallskip}\hline
\end{tabular}
\end{table}

\begin{figure*}
\begin{center}
  \includegraphics[width=0.3\textwidth]{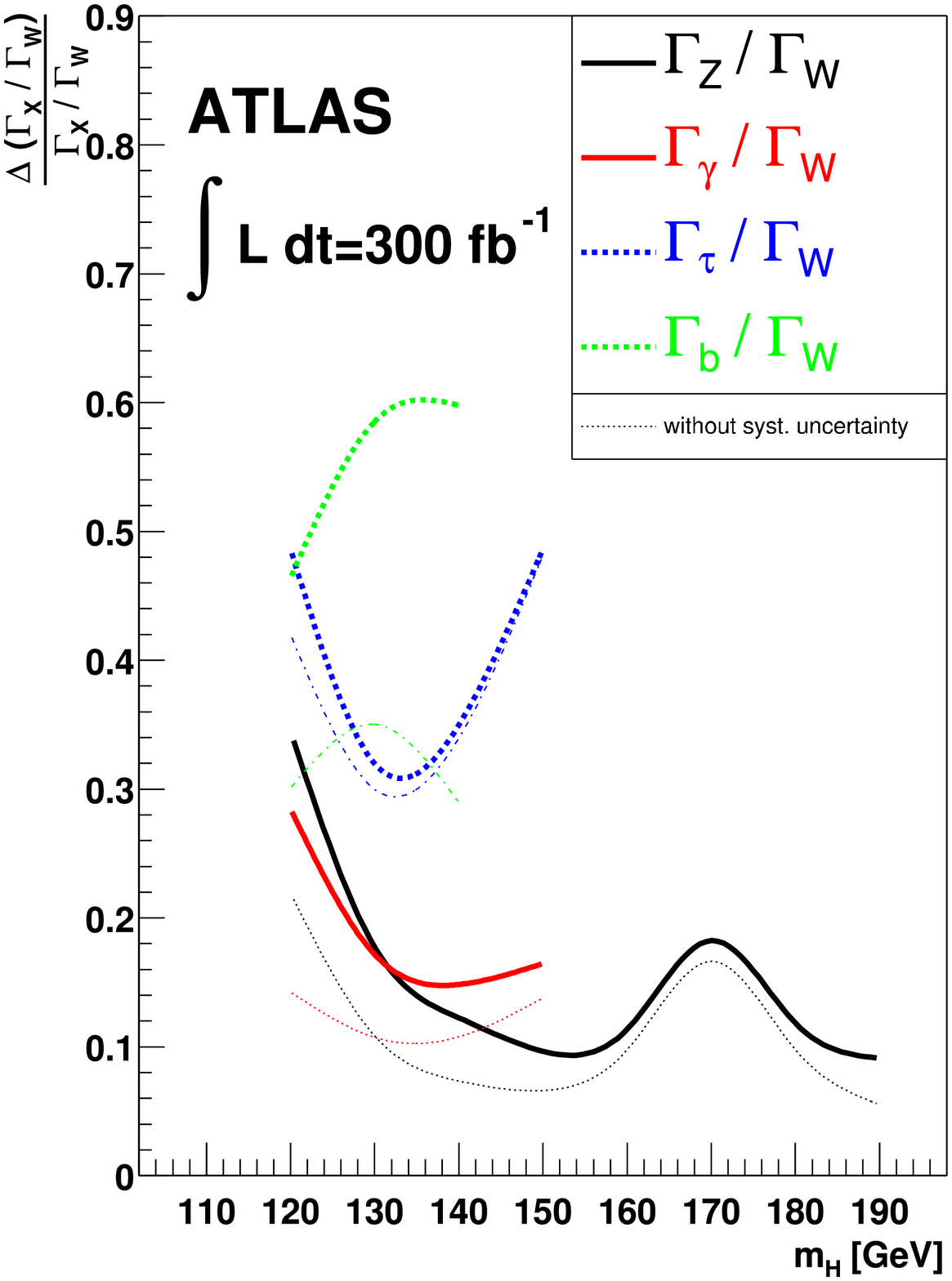}
  \includegraphics[width=0.3\textwidth]{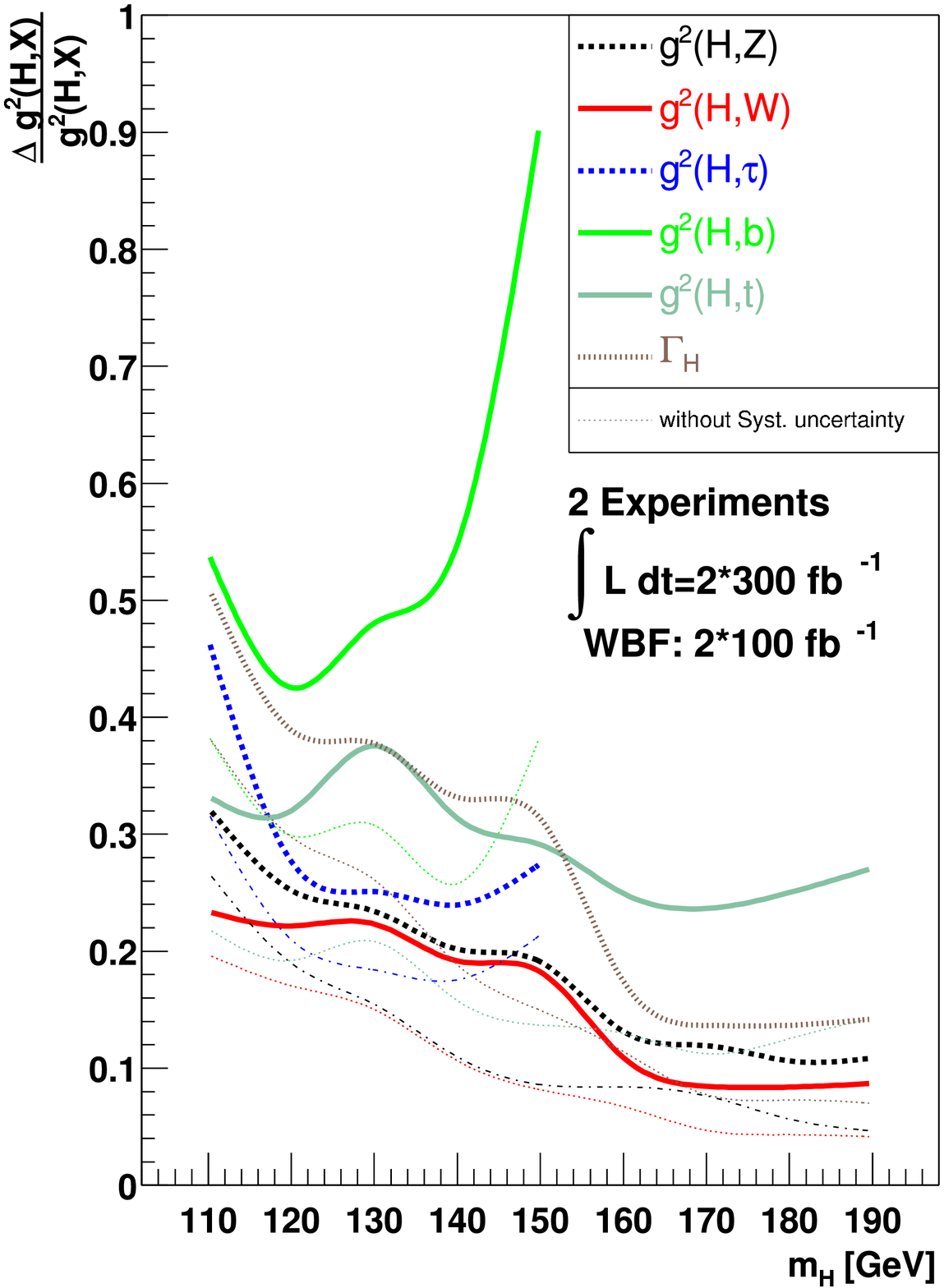}
  \includegraphics[width=0.3\textwidth]{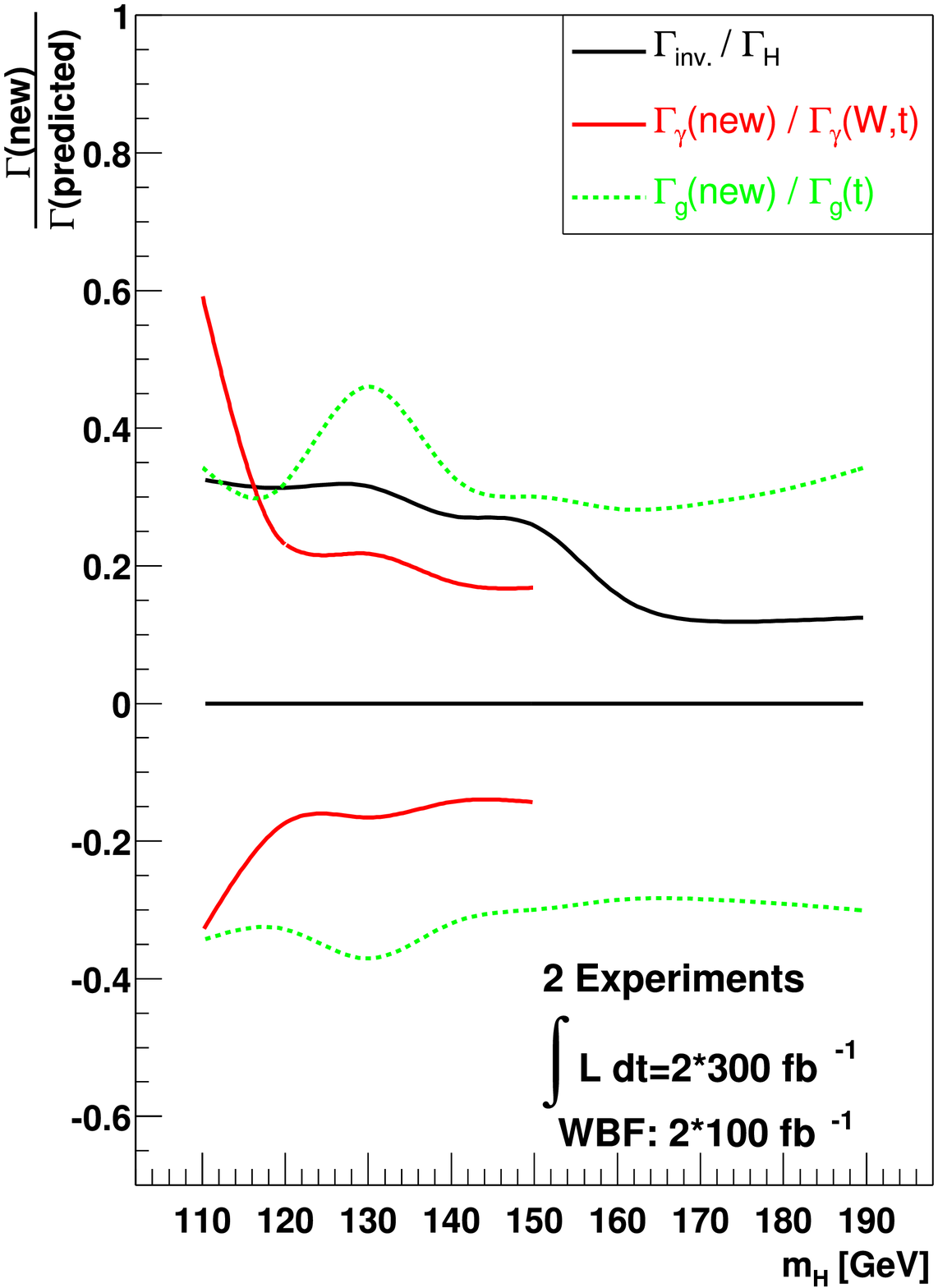}
\end{center}
\caption{Expected precision of the measurement of relative partial widths \cite{DUEHRSSEN_ATLAS} (\emph{left})
  using a likelihood fit under the assumption that only one boson contributes to the observed signal (only
  30\,fb$^{-1}$ were assumed for the weak boson fusion channels). Expected precision of the measurement of
  absolute couplings and the width of the boson (\emph{middle}) and expected precision of the measurement of
  additional partial widths \cite{DUEHRSSEN_PRD} (\emph{right}) using a likelihood fit under the additional
  assumption $\mathrm{\Gamma}_{V} < \mathrm{\Gamma}_{V}^{\mathrm{SM}}$.}
\label{fig:couplings_fit}
\end{figure*}

Since the total width of the boson in the considered mass range is not known it is not possible to extract the
partial widths and coupling constants from the observed event rates in a completely model-independent
way. Under the assumption that only one boson contributes to the observed signal it is possible to perform a
fit of ratios of partial widths. The results are shown in Fig.\ref{fig:couplings_fit} for an integrated
luminosity of 300\,fb$^{-1}$. For the weak boson fusion channels only 30\,fb$^{-1}$ were assumed because the
analyses of weak boson fusion rely on the reconstruction of jets in directions close to the beam axis and a
veto against additional jets in the central part of the detector, both of which have only been studied in
detail for the initial low-luminosity phase of LHC operation for which 30\,fb$^{-1}$ are expected. The ratios
are taken with respect to the partial width $\mathrm{\Gamma}_{W}$ because the $H \to WW$ channel has a
relatively high precision over the whole mass range used in the fit. The expected precisions for the ratios
$\mathrm{\Gamma}_{Z} / \mathrm{\Gamma}_{W}$ and $\mathrm{\Gamma}_{\gamma} / \mathrm{\Gamma}_{W}$ lie between
10\% and 35\%. For $\mathrm{\Gamma}_{\tau} / \mathrm{\Gamma}_{W}$ and $\mathrm{\Gamma}_{b} /
\mathrm{\Gamma}_{W}$ the relative uncertainties lie between 30\% and 60\%.

By making the additional assumption that the partial widths to weak bosons are not larger than in the Standard
Model ($\mathrm{\Gamma}_{V} < \mathrm{\Gamma}_{V}^{\mathrm{SM}}$), which is justified in any model containing
only Higgs doublets and singlets, the Higgs boson width can be constrained from above. The width is also
constrained from below by the likelihood \cite{DUEHRSSEN_ATLAS}, permitting a fit of the couplings and the
width. The results are shown in Fig.\ref{fig:couplings_fit}. The expected precisions are typically around 20\%
- 30\%. With the exception of the Higgs boson coupling to $b$ quarks the expected precisions are all better
than 50\%. The fit also allows for a determination of new contributions to the loops describing
$H\gamma\gamma$ and $Hgg$ couplings and an invisible width of the Higgs boson by including $\Gamma_{\gamma}$,
$\Gamma_{g}$ and a new $\Gamma_{inv}$ in the fit. The results for these contributions are also shown in
Fig.\ref{fig:couplings_fit}. The expected precisions are comparable to those for the couplings.

The likelihood function used for the fit can be used to calculate the expected discrepancy of the fit results
for a Standard Model Higgs boson from the expectations in the Minimal Supersymmetric Standard Model
(MSSM). Results for the expected exclusion of regions in the MSSM parameter space for several benchmark
scenarios are shown in \cite{DUEHRSSEN_PRD}.

\section{Determination of spin and CP quantum numbers}
\label{sec:spin_cp}

The Landau-Yang theorem \cite{yang} implies that the spin-1 hypothesis for the Higgs boson can be excluded if
the decay $H \to \gamma\gamma$ is observed.

\begin{figure}
\begin{center}
  \includegraphics[width=0.47\textwidth]{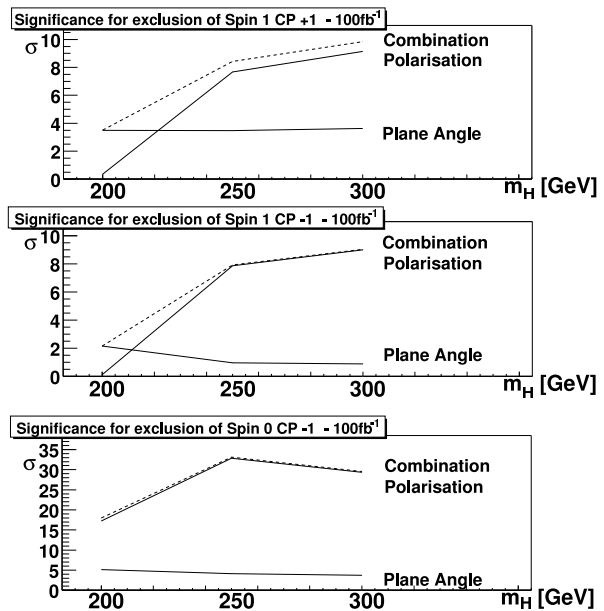}
\end{center}
\caption{Expected deviation of fit results for different combinations of spin and CP quantum numbers of the
 Higgs boson from the Standard Model expectation divided by the uncertainty expected for the Standard Model
 case \cite{buszello}.}
\label{fig:spin_cp_sig}
\end{figure}

The spin and CP quantum numbers of the Higgs boson can be studied in the channel $H \to ZZ \to 4l$
\cite{buszello,CMS_TDR} by using the angular distributions of the decay leptons. The sensitive observables
that are used are the angle $\phi$ between the planes spanned by the two lepton pairs and the angle $\theta$
between the momentum of a negatively charged lepton in the rest frame of the corresponding $Z$ boson and the
momentum of that $Z$ boson in the rest frame of the Higgs boson. The prospects for the exclusion of
non-Standard Model combinations of spin and CP quantum numbers with the ATLAS experiment have been studied
\cite{buszello} by performing fits of parametrizations to the angular distributions. The resulting expected
deviation from the Standard Model expectation divided by the expected error for the Standard Model case is
given as the significance in Fig.\ref{fig:spin_cp_sig}. The angle $\theta$ (``Polarisation'') provides a good
discrimination against the scalar CP odd case and, for masses larger than about 230\,GeV, against the spin-1
cases. For smaller masses near 200\,GeV the angle $\phi$ (``Plane Angle'') contributes significantly to the
combined significance.

\begin{figure*}
\begin{center}
  \includegraphics[width=0.32\textwidth]{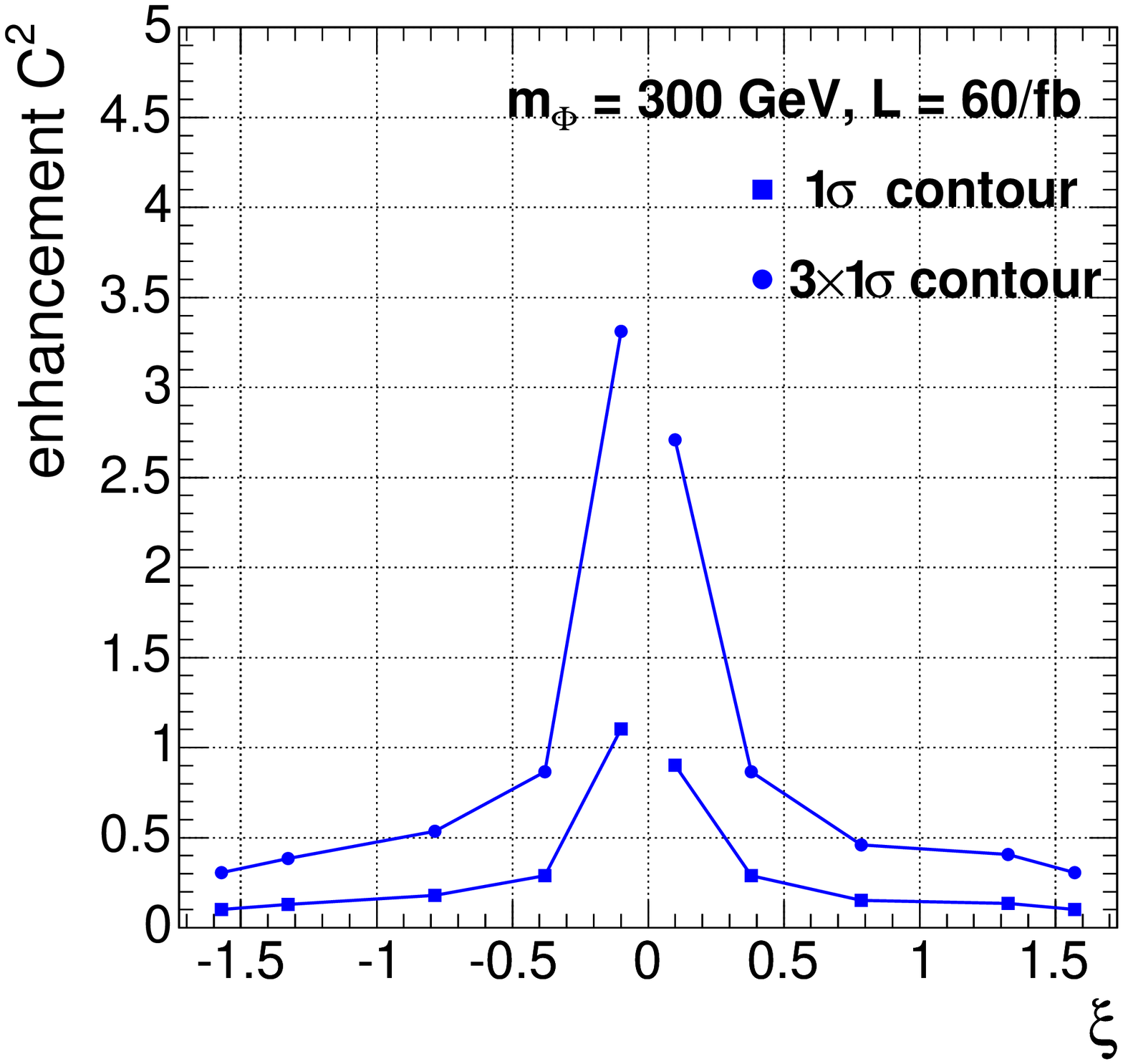}
  \includegraphics[width=0.32\textwidth, height=0.3\textwidth]{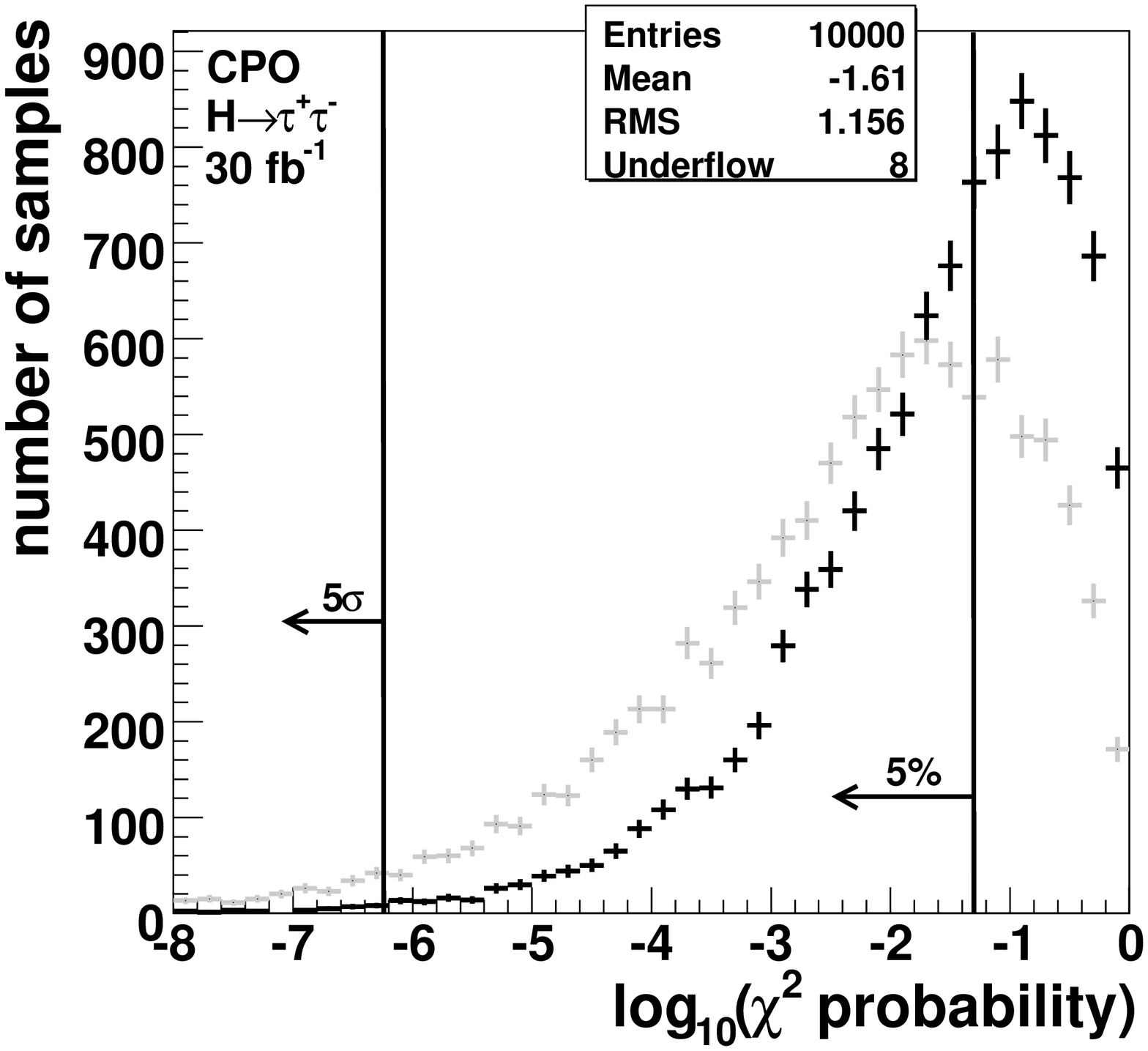}
  \includegraphics[width=0.32\textwidth, height=0.3\textwidth]{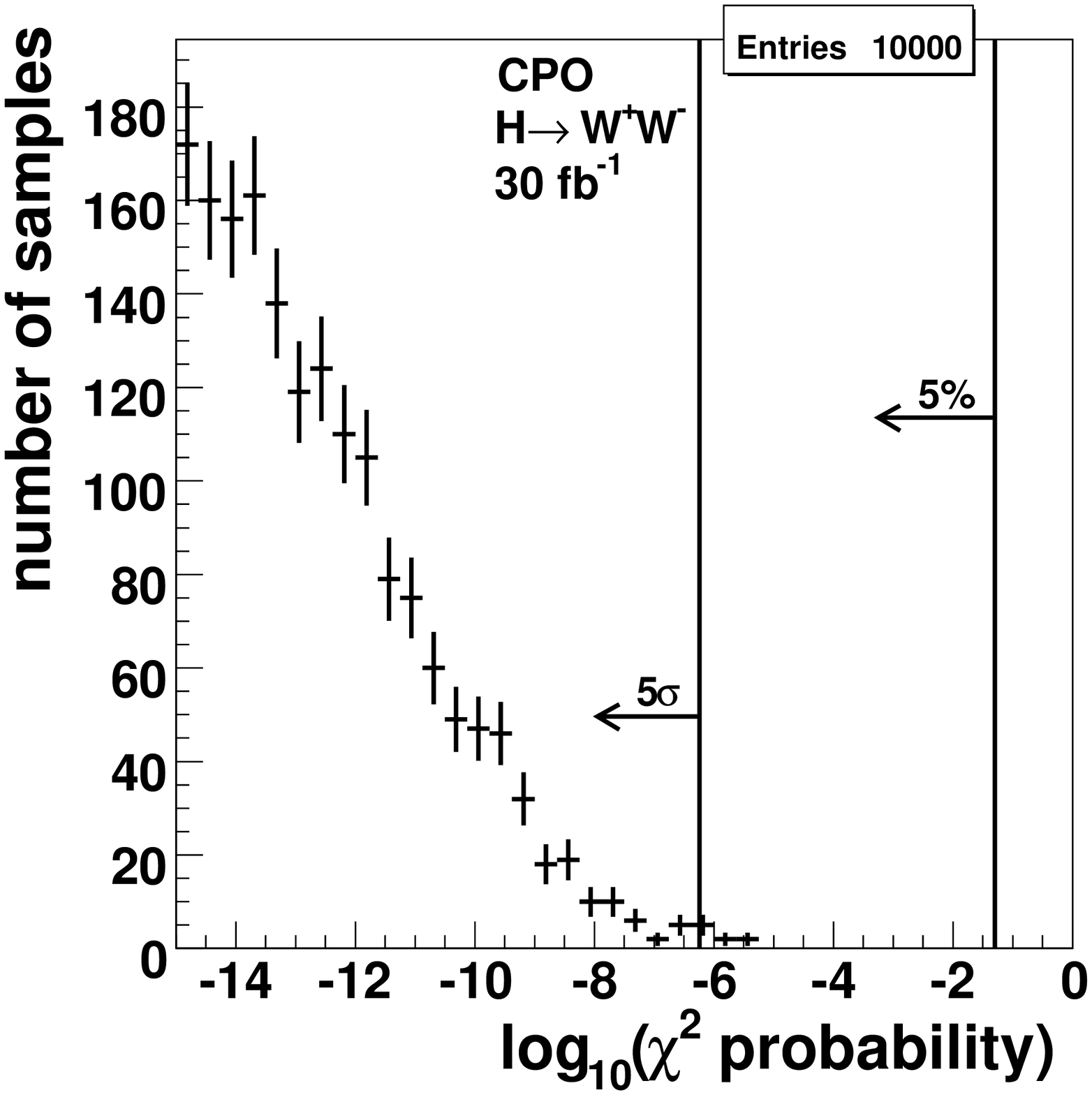}
\end{center}
\caption{Enhancement factor for the cross section required for the exclusion of the Standard Model hypothesis
  versus the parameter $\xi$ (see text) \cite{CMS_TDR} (\emph{left}). Distributions, obtained by performing a
  series of pseudo-experiments, of the $\chi^{2}$ probability of the hypothesis of a purely CP odd effective
  coupling for a Higgs boson mass of 120\,GeV (\emph{middle}) and for a Higgs boson mass of 160\,GeV
  \cite{ruwiedel} (\emph{right}). The gray distribution is obtained if backgrounds are neglected.}
\label{fig:hvv_structure}
\end{figure*}

\section{Determination of the structure of HVV couplings}

The most general coupling of a spin-0 boson to vector bosons contains, in addition to the Standard Model
coupling which is proportional to the metric tensor $g^{\mu\nu}$, a CP even and a CP odd term proportional to
combinations of the momenta of the participating particles. In the Standard Model the additional terms are
zero at tree level. They can be introduced by adding dimension-5 terms to the Lagrangian, examples being the
Standard Model effective $Hgg$ and $H\gamma \gamma$ couplings.

The prospects for the measurement of a contribution of $\tan \xi / m_{Z}^{2}$ times the CP odd effective
coupling in addition to the Standard Model coupling were studied \cite{CMS_TDR} for the CMS experiment. In the
study the angular distributions of the leptons in the channel $H \to ZZ \to 4l$ as described in
Sec.\ref{sec:spin_cp} were used. For the case of a Higgs boson mass of 300\,GeV and assuming an integrated
luminosity of 60\,fb$^{-1}$ the enhancement factor for the cross section required for the exclusion of the
Standard Model hypothesis for a given value of the parameter $\xi$ is shown in Fig.\ref{fig:hvv_structure}.

The structure of the coupling of the Higgs boson to vector bosons can also be studied in the weak boson fusion
process. The sensitive observable that is used is the difference $\Delta\phi_{jj}$ between the azimuthal
angles of the tagging jets which originate from the scattered quarks. The prospects for the exclusion of the
hypotheses of a purely CP even or a purely CP odd effective coupling for a Standard Model Higgs boson with the
ATLAS experiment were studied \cite{ruwiedel} by performing a $\chi^{2}$ test in the distribution of
$\Delta\phi_{jj}$. The results of the hypothesis test for a series of pseudo-experiments for the hypothesis of
a purely CP odd effective coupling and for a Higgs boson mass of 120\,GeV in the decay channel $H \to
\tau\tau$ and a Higgs boson mass of 160\,GeV in the decay channel $H \to WW$, assuming an integrated
luminosity of 30\,fb$^{-1}$ in each case, are shown in Fig.\ref{fig:hvv_structure}. For the case of a Higgs
boson mass of 120\,GeV an exclusion at a 95\% confidence level is observed in more than half of the
pseudo-experiments. For the case of a Higgs boson mass of 160\,GeV the observed $\chi^{2}$ probability almost
always corresponds to more than 5\,$\sigma$.

The prospects for the measurement of a contribution by the CP even effective coupling in addition to the
Standard Model coupling were studied \cite{ruwiedel} by performing a likelihood fit to the $\Delta\phi_{jj}$
distribution. This fit is sensitive to the interference between the effective coupling and the Standard Model
coupling. In a normalization such that a coupling of 1 reproduces the Standard Model cross section for a
purely effective coupling the expected standard deviation for the coupling from the fit, assuming an
integrated luminosity of 30\,fb$^{-1}$, is 0.20 for a Higgs boson mass of 120\,GeV using the $H \to \tau\tau$
channel and 0.09 for a Higgs boson mass of 160\,GeV using the $H \to WW \to ll\nu\nu$ channel.

\section{Summary}

The ATLAS and CMS experiments offer not only a good discovery potential for the Higgs boson but also promising
pros\-pects for the determination of its properties. The measurement of the Higgs boson mass is expected to
reach a precision around 0.1\% over a large range of masses. The width is expected to be measurable with a
precision better than 10\% for masses above 250\,GeV. Higgs boson couplings and width are expected to be
measurable under mild theoretical assumptions with typical precisions of a few 10\% in the mass range below
200\,GeV. Non-Standard Model combinations of Spin and CP quantum numbers are expected to result in significant
deviations from Standard Model expectations for Higgs masses above 200\,GeV. The vector boson fusion channel
offers good prospects for the study of CP properties of the Higgs boson and the structure of Higgs-vector
boson couplings, also in the range of Higgs masses below 200\,GeV.

\end{document}